\documentclass[twoside,a4paper,11pt]{sea10}
\usepackage{graphicx}
\usepackage{hyperref}
\usepackage{movie15}
\begin{document}
\pagenumbering{arabic}
\pagestyle{myheadings}
\thispagestyle{empty}
{\flushleft\includegraphics[width=\textwidth,bb=58 650 590 680]{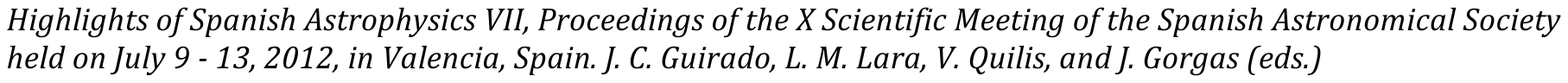}}
\vspace*{0.2cm}
\begin{flushleft}
{\bf {\LARGE
%
3D view on Virgo and field dwarf elliptical galaxies: late-type origin and environmental transformations
%
}\\
\vspace*{1cm}
%
Agnieszka Ry\'{s}$^{1,2}$,
Jes\'{u}s Falc\'{o}n-Barroso$^{1,2}$, 
and 
Glenn van de Ven$^{3}$
%
}\\
\vspace*{0.5cm}
%
$^{1}$
Instituto de Astrof\'{i}sica de Canarias, 38200 La Laguna, Tenerife, Spain\\
$^{2}$
Departamento de Astrof\'{i}sica, Universidad de La Laguna, 38205 La Laguna, Tenerife, Spain \\
$^{3}$
Max Planck Institute for Astronomy, K\"{o}nigstuhl 17, 69117 Heidelberg, Germany
%
\end{flushleft}
%
\markboth{
3D view on Virgo and field dwarf elliptical galaxies
}{ 
%
Agnieszka Ry\'{s} et al.
%
}
\thispagestyle{empty}
\vspace*{0.4cm}
\begin{minipage}[l]{0.09\textwidth}
\ 
\end{minipage}
\begin{minipage}[r]{0.9\textwidth}
\vspace{1cm}
\section*{Abstract}{\small

We show the effects of environmental evolution on Virgo cluster and field dwarf elliptical galaxies (dEs), presenting the first large-scale integral-field spectroscopic data for this galaxy class. The great variety of morphological, kinematic, and stellar population parameters seen in our data supports the claim that dEs are defunct dwarf spiral/irregular galaxies and points to a formation scenario that allows for a stochastic shaping of galaxy properties. We further investigate the properties of our sample by analyzing its kinematic and dynamical properties. We compare the level of rotational support of dEs and giant early-type galaxies and show that the properties of the former largely resemble those of giant fast-rotators. Based on our data, no trend exists between the level of rotational support in dEs and their location in the cluster. However, a tentative trend is seen in dark matter fraction: it increases for larger Virgocentric distances.
%
\normalsize}
\end{minipage}
%
%
%
\section{Introduction \label{intro}}

The dwarf elliptical galaxy class has been a point of much interest in recent years. Photometric surveys have revealed a whole range of substructures: spiral arms, bars, disks, twists, lenses (e.g. \cite{jerjen:2000}, \cite{barazza:2002}, \cite{graham:2003}, \cite{derijcke:2003}, \cite{lisker:2007}, \cite{janz:2012}). Through the  kinematic studies of, e.g. \cite{geha:2003}, \cite{chilingarian:2009}, \cite{toloba:2011} we have learned about the variety of rotation patterns, and stellar populations analyses have shown a variety of ages, metallicities, and their gradients (e.g. \cite{michielsen:2008}, \cite{paudel:2011}, \cite{koleva:2011}). Thus, dEs have come a long way from being considered simple low-mass extension of giant ellipticals to having their complicated structure and history unveiled in numerous studies which led to the conclusion that they are most likely environmentally-transformed late-type galaxies (see e.g. the review in \cite{kormendy:2012}). 

Since the effect of such transformation processes depends on the local environment density, we would expect galaxy properties to change with clustrocentric distance. Indeed, such trends have been found for the shape and the degree of substructure in Virgo (\cite{lisker:2007}), and for the ages and metallicities of both Virgo (\cite{michielsen:2008}) and Coma dEs (\cite{smith:2009}). With our integral-field spectroscopic data for Virgo dEs we have now turned to investigating their dynamical properties and the connection thereof to the location in the cluster.

\section{Data \& methods}

\begin{figure}
\center~
\includegraphics[width=15.0cm]{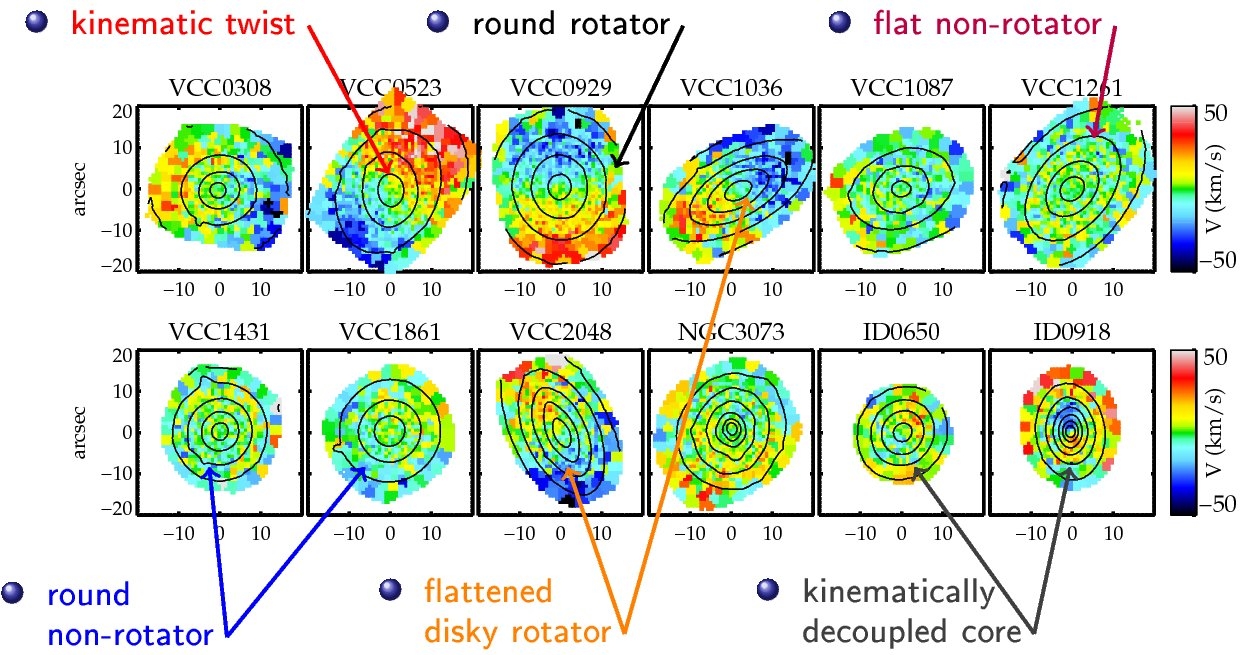} 
\caption{\label{fig1} Stellar velocity \textit{V} maps of our full sample shown here together to emphasize the kinematic variety. The level of rotation is not tied to flattening, we see kinematics twists (VCC\,0523, ID\,0918) and kinematically-decoupled components (ID\,0650 and ID\,0918). A full set of maps and profiles (\textit{V}, $\sigma$, and their errors) can be found in \cite{rys:2012}. 
}
\end{figure}

Details on data selection, observations and data reduction are presented in \cite{rys:2012}. In short, we observed 12 dEs: 9 in the Virgo Cluster and 3 in the field. Our objects span a wide range of ellipticities and distances from the cluster's center. The observations were carried out in Jan 2010 and Apr 2011 (8 nights in total) using the WHT/SAURON instrument at the Roque de los Muchachos Observatory in La Palma, with each galaxy typically exposed for 5h. For the extraction and calibration of the data we followed the procedures described in \cite{bacon:2001} using the specifically designed XSAURON software developed at the Centre de Recherche Astrophysique de Lyon (CRAL). The data were spatially binned to achieve the required minimum signal-to-noise ratio of 30 per pixel. The stellar absorption-line kinematics were derived for each galaxy by directly fitting the spectra in the pixel space using the penalized pixel-fitting method (pPXF) of \cite{cappellari:2004}.

\section{Kinematic variety}

These velocity maps shown in Figure~\ref{fig1} are the best example of the level of variety we are finding.  The maps show varying degrees of rotation: from $V_{max}\approx$40 km/s for VCC\,0523 to virtually no rotation for VCC\,1261. Rotation seems to be uncorrelated with flattening since we see all possible combinations of both. In addition to that, in two galaxies (ID\,0650 and ID\,0918) we observe large-scale kinematically-decoupled components, counter-rotating with regard to the main bodies. Also, kinematic twists are seen in VCC\,0523 and ID\,0918.

To relate this variety to the formation histories of these galaxies, we need to tie the results to the transformation mechanisms acting in cluster. The combined influence of ram-pressure stripping and harassment is the most likely answer, given that they allow for the random shaping of galaxy properties, based on the variety initial infall parameters (disk inclination and orbit type) and the number of encounters with large galaxies.

\section{Dynamical properties}
\subsection{Rotational support: dwarf vs. giant early-type galaxies}

\begin{figure}
\center~
\includegraphics[width=7.3cm]{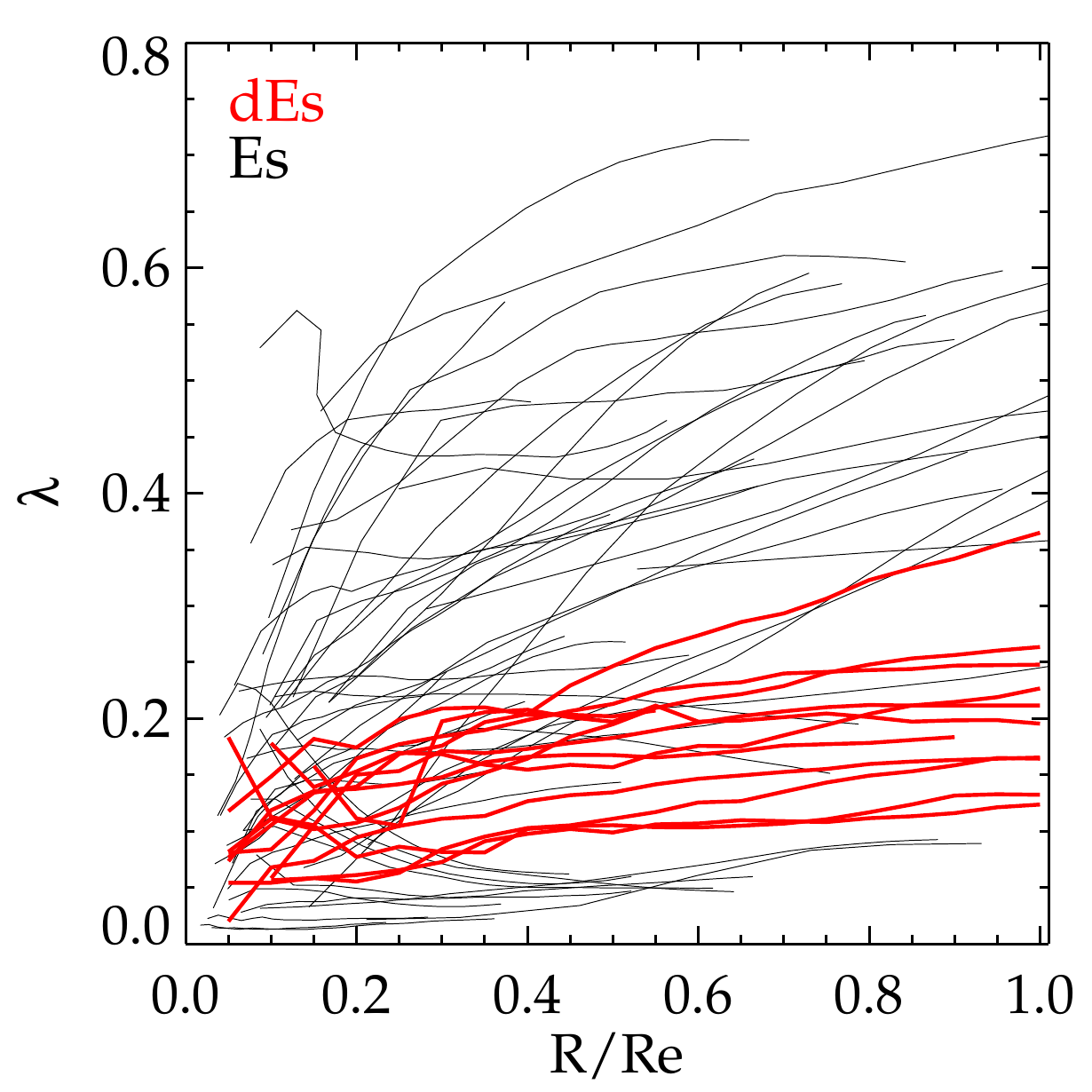} \hspace{0.4cm}
\includegraphics[width=7.3cm]{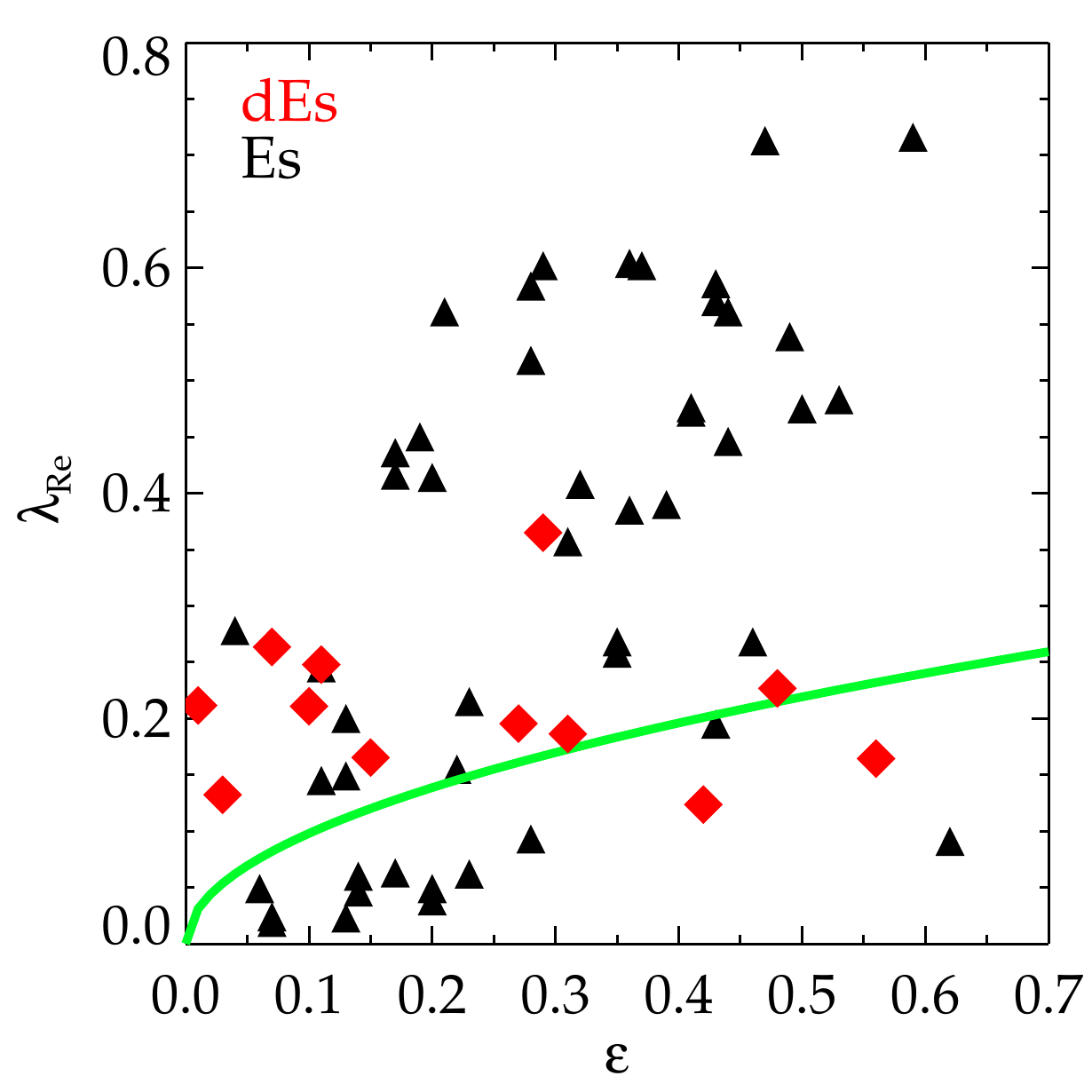} 
\caption{\label{fig2} \textit{Left}: Radial $\lambda_R$ profiles for the combined sample of dEs from this study (thick red lines) and ETGs of \cite{emsellem:2007} (thin grey lines). Slow-rotating ETGs have profiles that are decreasing or nearly flat and fast-rotators have increasing profiles. \textit{Right}: $\lambda_R$ versus ellipticity for the same sample (dEs -- red diamonds, Es -- black triangles). The green solid line corresponds to $0.31\cdot \sqrt \epsilon$, the threshold separating the fast- and slow-rotator families defined by \cite{emsellem:2011})
}
\end{figure}

Since dEs used to be thought of as a low-luminosity extension of giant ellipticals (Es), we present here a comparison of the kinematic properties of our dEs and early-type giants (ETGs) of \cite{emsellem:2007}. To estimate the amount of rotation in our objects we have therefore decided to use the new $\lambda_R$ parameter defined in \cite{emsellem:2007} (Figure~\ref{fig2}). $\lambda_R$ is better suited for angular momentum estimate than the traditionally employed V/$\sigma$.

For the most part dEs behave like the members of the fast-rotator family: their integrated $\lambda_{Re}$ values are above or on the line that divides the ETG group into slow and fast rotators. Their $\lambda_R$ profiles also tend to increase for larger R/R$_e$ or lie in between those of slow- and fast-rotator groups. A direct comparison is, however, tricky.  A range of differences stand behind each type that include photometric, kinematic, dynamic, and X-ray properties (see, e.g., \cite{emsellem:2007}, \cite{emsellem:2011}). Thus, the empirical dividing line cannot be easily used to accommodate dEs in the same diagram. Moreover, by their very nature the data for dEs are noisier, which will produce a positive offset in the calculated $\lambda_R$. One, thus, needs to be cautious when devising their analysis approach in such cases as ours. 

Nevertheless, these results support the claim of dEs having late-type origin. Giant fast rotators (most of which are S0s) have properties more similar to those of late-type galaxies. S0s are believed to be faded spirals (e.g. \cite{kormendy:2012}), so the fact that our dEs show a similar level of variety as the SAURON survey's fast rotators only strengthens the argument in favor of them being transformed late-type galaxies.

\subsection{Global trends: rotational support and dark matter fraction}

\begin{figure}
\center~
\includegraphics[width=7.3cm]{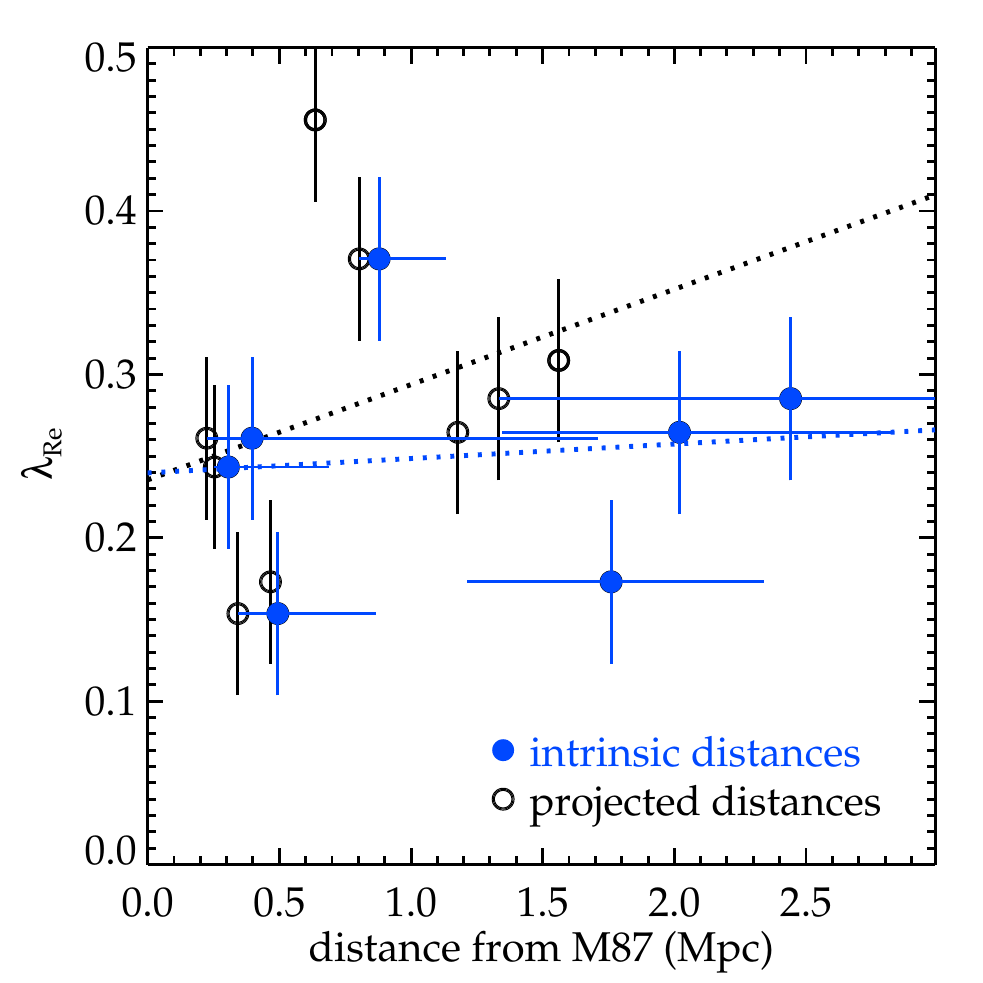} \hspace{0.4cm}
\includegraphics[width=7.3cm]{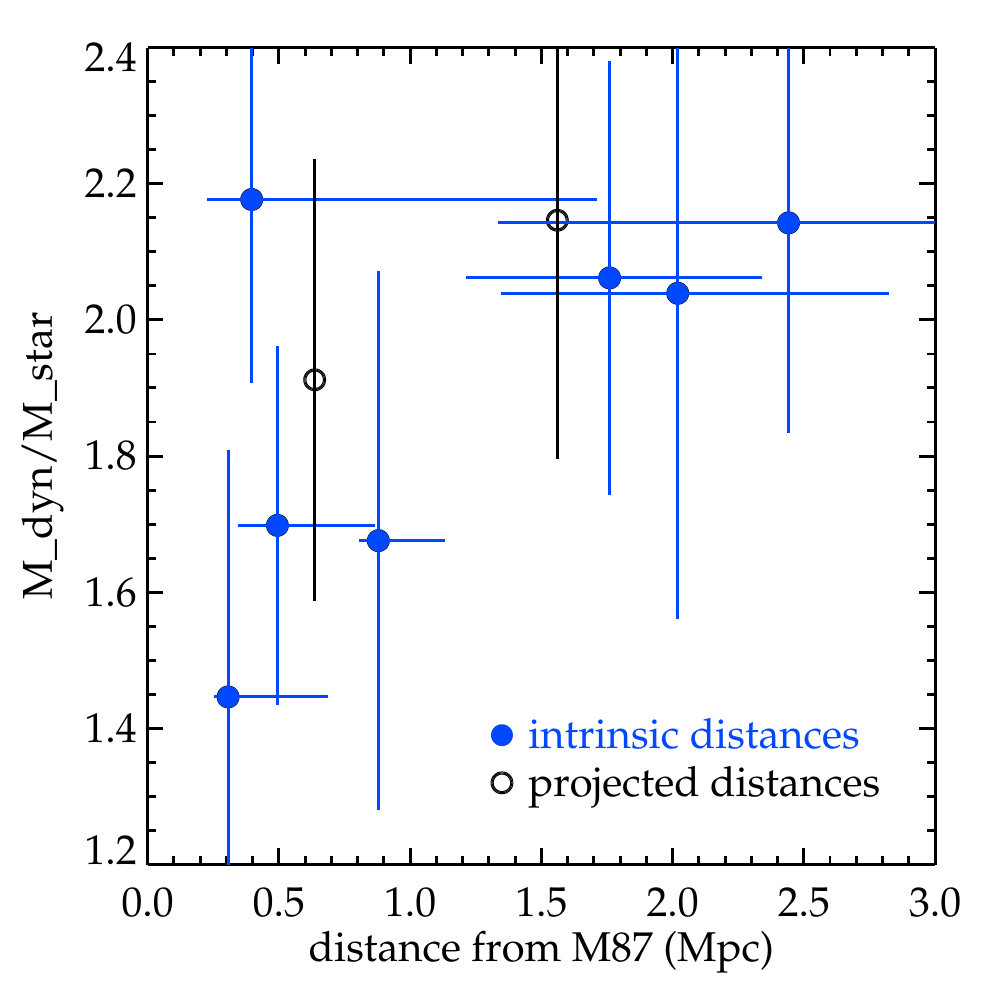} 
\caption{\label{fig3} \textit{Left}: specific angular momentum $vs$ Virgocentric distance. For all galaxies both intrinsic and projected  distances are shown. No radial trend is found when intrinsic distances are used (\cite{rys:2012}). \textit{Right}: dark-to-stellar matter ratio as a function of Virgocentric distance. Intrinsic distances are plotted where available, otherwise projected distances are provided. A tentative trend is seen: the ratio increases with increasing distance (Ry\'{s} et al., in prep.)
}
\end{figure}

As we have already discussed in \cite{rys:2012}, the analysis of any property trends with clustrocentric distance, should be performed using intrinsic (i.e. deprojected) distances in order to minimize the projection bias when dealing with incomplete samples. We show this effect in the left panel of Figure~\ref{fig3} where a trend of rotational support versus Virgocentric distance is shown for both projected and intrinsic distances. Our results, albeit based on a limited-size sample, tentatively show little if any relation between the two aforementioned quantities.

We have also investigated the relation between the dark matter fraction and the location in the cluster, motivated by the simulations of \cite{moore:1998} and \cite{smith:2010} who suggest that (at a fixed radius) dark matter might be lost more easily due to its different orbital distribution. The preliminary results of this analysis are shown in the right panel of Figure~\ref{fig3}. The total (dark+stellar) masses come from Jeans axisymmetric MGE models \cite{cappellari:2008} which we constructed combining our SAURON stellar kinematics and SDSS r-band imaging. To obtain the stellar mass estimates we used the \cite{lisker:2008} SDSS g-r colors and the color-mass conversion of \cite{bell:2003}

\section{Discussion and conclusions}

The variety of properties we find does not mean that we should necessarily look for separate formation mechanisms for each dE subclass. The environmental reshaping processes are to a certain degree random processes that are able to produce different products from similar input objects. Therefore, an effort has to be made to create a big picture and build a fully consistent transformation scenario. 

The results presented here are only indicative of possible trends, given the sample size and, in the case of dark matter trend, the uncertainties on the mass estimates. Nevertheless, studying a small number of galaxies in great detail has (in fact, always) been a necessary first step, as it is in this way that tentative trends are found that can later be investigated with larger samples. It is also a good starting point for a discussion on what is actually expected in terms of trends from a theoretical point of view.

What is important is the predicted timescales of transformation events, and their comparison with the average cluster crossing times. This information makes it possible to determine whether a given scenario is properly predicting the fate of infalling galaxies or not. If the different timescales do not add up, then the simulations need to be revised and/or perhaps other scenarios considered. 

Drawing conclusions from comparing the different timescales is not a straightforward task. To name a few problems, galaxies' orbits in clusters differ from one another: at the same distance form the center one galaxy might be approaching the center on its first pass while another will have travelled across the cluster already more than once. We do not know how long it has been since galaxies fell into a cluster, we are also limited in our analysis by projection effects: galaxies appearing to be in the center might in fact be located at any clustrocentric distance. 

Thus, in order to try to tackle the obstacles discussed above and have a chance at disentangling the importance of different processes, a holistic approach needs to be applied that combines dynamical and stellar populations studies of large data samples. With this approach in mind, we hope not only to give a qualitative assessment of various formation mechanisms, but also to disentangle the relative strength and contributions of each of them, thus quantitatively determining the actual transformation paths for these galaxies.

%
\small  
%
\section*{Acknowledgments}   
AR would like to thank the Spanish Astronomical Society for financial assistance to attend the society's $10^{th}$ Scientific Meeting in Valencia in July 2012 where this work was presented. JFB acknowledges support from the Ram\'{o}n y Cajal Program financed by the Spanish Ministry of Economy and Competitiveness (MINECO). This research has been supported by the Spanish Ministry of Economy and Competitiveness (MINECO) under grants AYA2010-21322-C03-02 and AIB-2010-DE-00227. AR acknowledges the repeated hospitality of the Max Planck Institute for Astronomy in Heidelberg, to which collaborative visits contributed to the quality of this work.

%

%
\end{document}